\documentclass[preprint,12pt]{elsarticle}

\usepackage{amssymb}
\usepackage{booktabs}
\usepackage{float}
\usepackage{xcolor}

\journal{Optical materials}

\begin{document}

\begin{frontmatter}



\title{Coherent control of nonreciprocal optical properties of the defect modes in 1D defective photonic crystals with atomic doping}



           %

\author[1]{Nancy Ghangas}
\ead{nancy.19phz0006@iitrpr.ac.in}
\author[1]{Shubhrangshu Dasgupta}
\affiliation[1]{organization={Department of physics},
            addressline={Indian Institute of Technology}, 
            city={Rupnagar},
            postcode={160001}, 
            state={Punjab},
            country={India}
            }
\begin{abstract}
We investigate the spectral properties of photonic crystals, lacking parity-time (PT) symmetry, using scattering matrix formalism. We show using the symmetry properties of matrices that a defective photonic crystal, doped with three-level atoms, breaks PT symmetry. For the two defect modes lying in the bandgap region, both the reflection and the absorption become nonreciprocal, while the transmission remains reciprocal. We show that the relevant energy spectra do not exhibit exceptional points, thereby invalidating its necessity to achieve nonreciprocity in reflection and absorption. We further demonstrate how this nonreciprocity can be coherently controlled using the driving field and dissipation rates of the atoms.
\end{abstract}




\end{frontmatter}


\section{Introduction}\label{sec1}
Photonic Crystals (PCs) with defects exhibit resonances in the band gap region. These resonances, known as defect modes, correspond to certain electromagnetic standing waves, akin to those in a Fabry-Perot resonator \cite{PhysRevLett.58.2059}. Such resonances appear when a phase slip (i.e., a defect layer) is introduced in a periodic configuration. The defective PCs therefore work as a cavity, with the defect modes as the corresponding cavity modes. Interestingly, in the presence of weak randomness in the periodicity, the field intensity gets enhanced at the defect site in the photonic band gap, leading to field localization \cite{PhysRevLett.58.2486}.
These PCs serve as valuable tools for fabricating high-quality narrowband frequency selective filters, sensors, and demultiplexers  \cite{PhysRevLett.58.2059, Article5, article6, article7}. In recent years, it has been shown that the optical properties and the quality factor of the PCs can be manipulated via band gap engineering using different types of defect layer materials \cite{article2, 10.1063/1.5089413, XU20155158, PhysRevA.101.043830,A1,A2,article8, FENG2005330, LIU200935,PhysRevB.57.3815}.
Furthermore, the analytical and numerical solutions for band structures in the presence of impurities have been elucidated and experimentally validated in \cite{Smith:93,Luna-Acosta_2008}.

Recently, Parity-Time (PT) symmetric systems in optical periodic structures have attracted much attention. In these structures, the Hamiltonian remains invariant upon parity (P) and time reversal (T). In the context of quantum mechanics, a Hamiltonian $\hat{H}=\frac{{\hat{p}}^2}{2m}+V(\hat{x})$ (where $\hat{x}$ and $\hat{p}$ are position and momentum operators respectively, $m$ is the mass, and $V(x)$ represents potential energy) is considered to be PT-symmetric, if $(PT)\hat{H}=\hat{H}(PT)$, (i.e., if $H$ and $PT$ commute with each other and $H$ shares common eigenvectors with the PT operator). 
Further, PT symmetry requires that the real part $V_R(x)$ of the complex potential is an even function of position $x$, whereas the imaginary part $V_I(x)$ is odd; that is, the Hamiltonian must have the form $\hat{H}=\frac{{\hat{p}}^2}{2m}+V_R(\hat{x})+\iota V_I(\hat{x})$. In optics, such PT-symmetric potentials can be realized through a combination of index guiding and gain/loss regions and the complex refractive index plays the role of complex optical potential as discussed and experimentally realized in \cite{ruter2010observation}.
Hence, PT symmetry manifests as a periodic modulation of refractive index or gain/loss distribution, leading to a non-Hermitian Hamiltonian and exhibiting certain extraordinary optical properties, including coherent perfect absorption (CPA) \cite{PhysRevA.101.063822}, and various unidirectional optical effects \cite{PhysRevLett.106.213901,cryst12020199, ZHANG2021110771, PhysRevLett.113.123004, Longhi_2015, Horsley2015, PhysRevLett.112.043904}. 
The CPA has also been exploited for multipurpose sensor applications, using Janus metasurface \cite{https://doi.org/10.1002/andp.202100499} and the PCs containing magnetized Yttrium Iron Garnet \cite{WU2023109422}, however in the terahertz regime. We further note that the direction dependence of absorption based on the electromagnetically induced absorption in a Silicon-analyte PC \cite{10287989} and combining epsilon-near-zero edge states in InSb material and photonic bandgap edge state of PCs \cite{10247234} has also been explored for biosensor application. However, this effect can be modulated by adjusting either the angle of incidence, the thickness of the metal, or the temperature. In contrast, we show how the unidirectional properties of the PC can be controlled using a coherent electromagnetic field incident parallel to the axis of the PC in the visible regime.

In this paper, we focus on the non-PT-symmetric defective PCs in which studies of such unidirectional and non-reciprocal properties are quite limited. As in the case of PT-symmetric PCs \cite{article14,PhysRevA.87.012103,Hang_2016}, these works primarily focused only on bilayer optical systems exhibiting singularities for a tailored set of gain/loss parameters  (e.g., as in plasmonic waveguide experiments \cite{Shen:14, article13,HuangShenMinFanVeronis+2017+977+996}).  %
Such singularities are associated with the PT-symmetry-breaking and phase transition at a specific frequency where either reflection or transmission diverges \cite{app10030823}.  
On the contrary, in this work, we show that a {\it multilayered} periodic optical system, with a defect layer doped with atoms, breaks PT-symmetry and exhibits non-reciprocal optical properties in the absence of any singularity. 
The non-reciprocity is primarily demonstrated when the transition frequency of atoms is in resonance with either of the two defect modes. The delicate balance of gain/loss parameters is no longer relevant. It is noteworthy that non-reciprocity entails reduced reflection and a near-complete absorption for a specific direction of propagation, and vice versa for the reverse propagation. While non-reciprocity has traditionally been explored within the framework of transmission, recent studies have expanded its scope to define it in terms of reflection and absorption \cite{Chaung:20,krichevtsov1993spontaneous,10287989}.
We emphasize that though the unidirectional optical effect has been reported in the literature, such a direction-dependence of these effects within the non PT-symmetric systems remains unaddressed so far. Typically, optical systems exhibit reciprocity except at exceptional points as reported. Our findings illustrate that non-reciprocity in reflection and absorption can be achieved {\it even if the exceptional point does not exist} while maintaining reciprocity in transmission.   

More importantly, it is to be noted that in the case of em waves, coherence indicates a consistent phase relationship that the em waves composing the light exhibit  as they propagate through space resulting in a well defined interference pattern. In our case, it is interference and coherence effects which leads to PBG formation. By controlling relative amplitude of control field, one can enhance or suppress specific pathways and transitions that can lead to the modification of refractive index of the atomic medium.
In this paper, we demonstrate a dynamical control of the optical properties of the PC, by using a strong coherent electromagnetic field. This field can be suitably tuned to an atomic transition within the defect, thereby leading to a coherent manipulation of the refractive index of the defect layer. Therefore, the non-reciprocal effect can be coherently controlled by manipulating just a single parameter - the amplitude of a resonant control field - unlike in \cite{Hang_2016} in which three parameters are to be adjusted to achieve the desired optical effect. At a critical value of the control field, the reflection nearly vanishes and the absorption coefficient approaches unity, indicating a critical coupling condition. This is unlike in the case of heterostructure PCs \cite{Zhang_2014, 2022FrP....1019214Y, Zhang:10,DEB20104764} in which one can achieve complete absorption, however, with enhanced reflection arising due to impedance mismatch at the interface of the defect layer. We emphasize that in our case, complete absorption refers to critical coupling {\it considering absorption in one direction at a time} rather than reported CPA phenomenon \cite{Hang_2016,PhysRevA.82.031801,article14}. Notably, coherent control of non-reciprocal reflections has been investigated in \cite{Chaung:20} with spatial modulation of the coupling in a cold Rubidium atomic lattice. Further, control of global PT-symmetry via local nonlinearity in lattices has also been explored experimentally \cite{nonlinearPTsymm}. However, these investigations primarily pertain to PT-symmetric, anti-PT-symmetric, and transition to non-PT-symmetric configurations via exceptional points. Our results substantially diverge from these works, both in the underlying physics and in the chosen system. 



When the defect layer is doped with atomic vapor, the scenario becomes similar to atoms embedded within an optical cavity. In our system, the PC acts as a two-mode nanocavity in the visible frequency range. We observe a spectral inversion in the reflection and absorption behaviour for the two modes, contingent upon the direction of incidence. Consequently, these modes can be denoted as non-PT dual pairs. 
in contrast to the above-mentioned similar works. Our analysis not only provides insight into the scattering properties of PC in PT-symmetry-broken regime, but also unveils the reliance of the asymmetric propagation on the control field and atomic decay rates which has not been reported earlier. 

In our analysis, we employ the standard transfer matrix and scattering matrix formalism \cite{doi:10.1080/00107510902888338}. The PC exhibits non-PT-symmetric behavior by violating the symmetry relations of the corresponding matrices. While transmission within the crystal remains reciprocal, reflection and absorption for two modes lying in the band gap of the crystal exhibit non-reciprocal behavior. 

This paper is organized as follows: In Section 2, we introduce the scattering matrix along with the relevant Hamiltonian of the PC. In Section 3, we present numerical results to demonstrate the non-reciprocal interplay of absorption and reflection at two defect modes and their coherent control using atomic doping.  Finally, we conclude the paper in the Section 4.
\section{Model and analysis}\label{sec2}
\begin{figure}[htb]
\centerline{
\includegraphics[width=8.8cm]{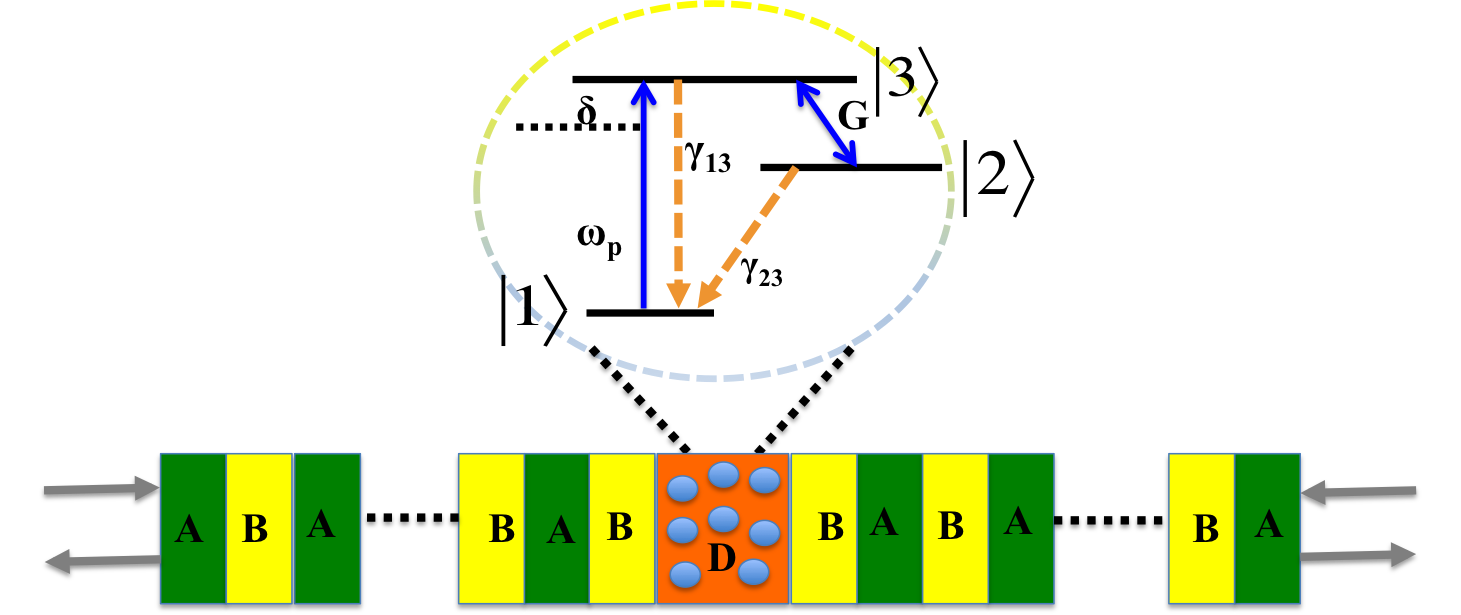}}
\caption{A schematic diagram showing the periodic layers of two different media A and B, with a defect layer D doped with atoms in three-level $\Lambda$-type configuration.}
\label{fig:1}
\end{figure}

We consider a periodic structure, shown in Fig.\ref{fig:1}, in which a quarter-wave stack with [$A(BA)^5BDB(AB)^5A$] configuration of alternate layers A and B, with respective refractive indices $n_A$ and $n_B$. The term `quarter-wave' corresponds to an optical thickness of these layers, given by $n_Ad_A=n_Bd_B=\lambda_0/4$, where $\lambda_0=632.8$ nm is a reference wavelength and $d_k$ is the physical thickness of the layer $k\in A, B$. A half-wave layer, D, $n_Dd_D=\lambda_0/2$ with refractive index $n_D$ is acting as a substrate layer doped with an atomic ensemble. In the following, we choose 
a three-level $\Lambda$-type configuration for doping, where a weak probe field with frequency $\omega_d$ drives the transition $|1\rangle \rightarrow |3\rangle$ and a control field with Rabi frequency $2G$ is applied to the $|2\rangle \rightarrow |3\rangle$ transition. The dielectric function $\epsilon(\omega)$ of the doped layer is given by
$\epsilon(\omega)=\epsilon_D+\chi(\omega)$ 
where, $\epsilon_D={n_D}^2$ is the dielectric constant of the substrate defect layer and $\chi(\omega)=\chi'+\iota\chi"$ is the susceptibility of the atoms in the defect. The real ($\chi'$) and imaginary parts ($\chi"$), representing dispersion and absorption respectively, are given by \cite{SAHRAI201366}
\begin{equation}\label{subeq:3}
\chi'=\frac{\delta[G^2-4(\gamma_{23}^2+\delta^2)]}{D'/4\alpha},\;
\chi"=\frac{4\gamma_{13}(\gamma_{23}^2+\delta^2)+\gamma_{23}G^2}{D'/4\alpha},
\end{equation}
where $\gamma_{ij}$ is the spontaneous decay rate in $|j\rangle\rightarrow |i\rangle$ transition, $\delta$ is the probe detuning with the defect mode, $\alpha=2N|d_{31}|^2\gamma_0/\epsilon_0\hbar$ 
is a constant, $N$ is the number density of atoms, $d_{31}$ is the dipole moment matrix element, $\gamma_0$ is a normalization constant, and $D'=16({\gamma_{23}}^2+\delta^2)({\gamma_{13}}^2+\delta^2)+8({\gamma_{13}}{\gamma_{23}}-\delta^2)G^2+G^4$. It is essential to note that these types of crystal configurations have been extensively realized experimentally in \cite{jmmp1010006,lagoudakis2013deterministically,RADISHEV2021118404} for various QED studies, although with involvement of some challenges in fabrication.

The input and output fields from both the left and the right sides of any $j$th layer ($j\in A, B, D$) are related to each other via the transfer matrix $T_j$, whose elements are
$T_{11}={T_{22}}^*= e^{-\iota\alpha_+}[A_+\cos x+\iota B_+\sin x] $ and $T_{12}=T_{21}^*= e^{\iota\alpha_-}[A_-\cos x+\iota B_-\sin x]$, with $x={k_jd_j}$, $\alpha_\pm=(k_L\pm k_R)d_j/2$, $A_\pm=(1\pm k_L/k_j)/2$, and $B_\pm=(k_j/k_R\pm k_L/k_j)/2$. Here $k_j={\sqrt{\epsilon_j(\omega)}}\omega/c$, is the wave vector amplitude inside the $j$th layer, $k_L$ and $k_R$ are that of the layers, on the left and the right, respectively, to this $j$th layer, and $c$ is the speed of light in vacuum.

The transfer matrix for a single layer is therefore $M_j=P_jT_j$, where $P_j= {\rm diag}[e^{\iota k_{j-1}d_j},e^{-\iota k_{j-1}d_j}]$ is the corresponding propagation matrix, referring to the phases acquired during propagation inside the layer in two opposite directions.  For the entire periodic structure (Fig. \ref{fig:1}), we then can write the transfer matrix as  
$M = M_{right}(M_BM_A)^5M_BM_DM_B(M_AM_B)^5M_{left}$.
The inputs and outputs of the periodic structures are often related to each other by a scattering matrix  $S=
   \left[\begin{array}{cc}
     r_f  & t_b \\
t_f& r_b
   \end{array}\right]$, via the relation $[b\;\; d]^T = S [a\;\; c]^T$, where $T$ represents transposition of a matrix. 
Here $t_f$ ($t_b$) and $r_f$ ($r_b$) are the transmission and the reflection coefficients, respectively, for the forward (backward) direction of propagation. These coefficients can also be expressed by the  matrix elements of $M$: 
$t_f = M_{11}-M_{12}M_{21}/M_{22}$, $r_f = -M_{21}/M_{22}$, 
$t_b = 1/M_{22}$, and $r_b = M_{12}/M_{22}$.
Clearly, the matrix $S$ above is not PT-symmetric \cite{PhysRevA.87.012103}, as $M_{11} \ne M_{22}^{*}$ and $M_{12} \ne -M_{21}^{*}$.
Note that the transfer matrix of a non-PT-symmetric system satisfies $M^{-1}\ne M^{*}$. In addition, we obtain the eigenvalues and eigenvectors of $S$ and the corresponding Hamiltonian of the PC. 
The Hamiltonian matrix $H$ can be related to the scattering matrix as $S = \exp({-\iota H})$ and $H = \iota\ln S$.

\begin{figure}[htb]
\centerline{
\includegraphics[width=8.8cm]{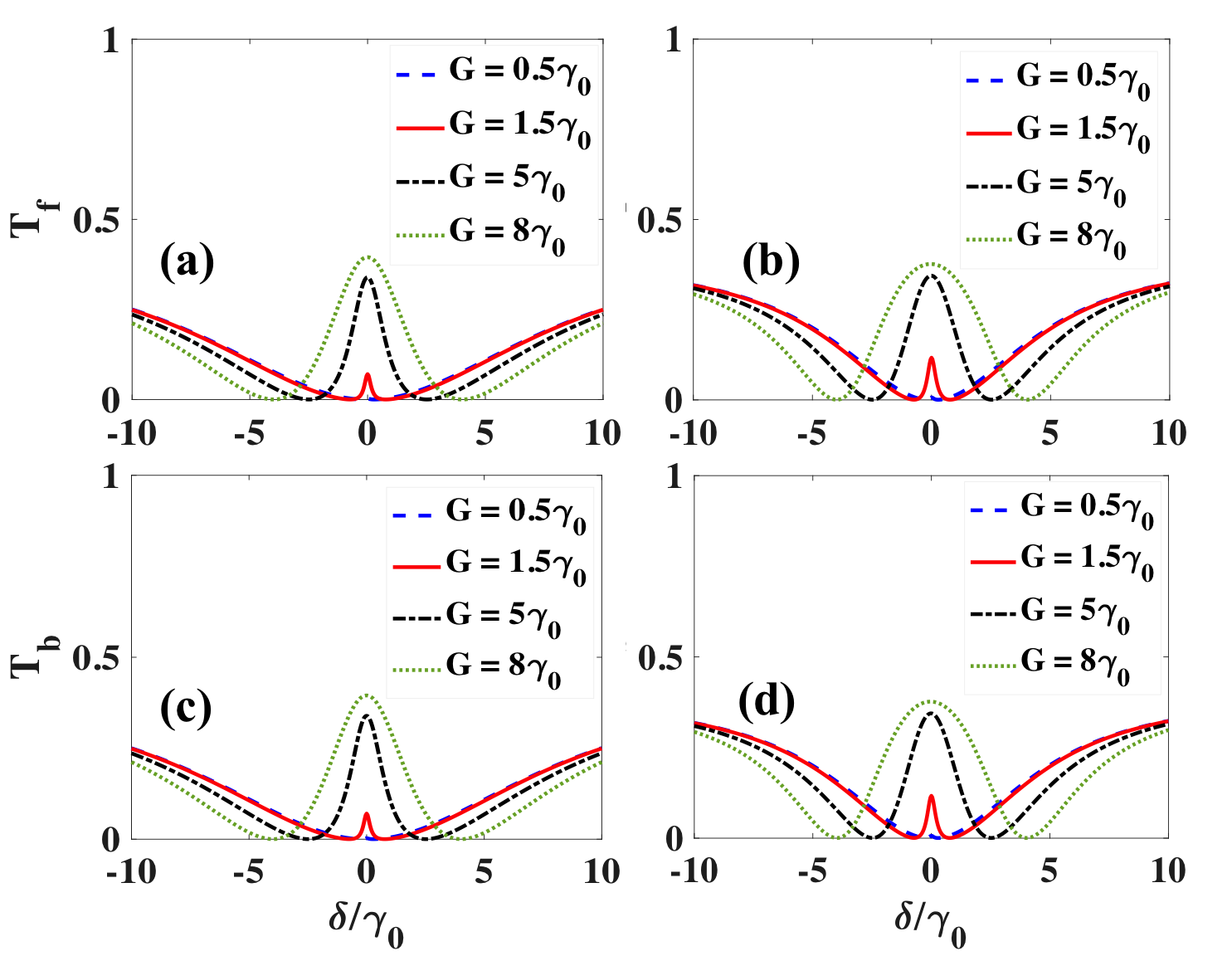}}
\caption{Transmission spectra [$T_{f,b}=|t_{f,b}|^2$] of the PC for different values of $G$ in (a),(b) forward direction and (c),(d) backward direction for two modes $\omega_1$ and $\omega_2$. We have chosen $\gamma_{12}=\gamma_{13}=\gamma_{23}=2\Gamma_{21}=0.1\gamma_0$, $\alpha=0.05\gamma_0$, and $\gamma_0=2\pi \times 9.978$ MHz.}
\label{fig2}
\end{figure}

\begin{table}
\centering
\caption{Parameters of the layers in the PC}
\label{tbl1}

\begin{tabular}{|l|c|c|}

\hline
  Refractive index of substrate layer  &  $n_D$  & 2.1 (Boron nitride)  \\
  
\hline
    Refractive index of defect layer &     $n_T$ &$\sqrt{\epsilon(\omega)}$  \\
    
\hline
    Width of defect layer D & $d_D$ & 150.6nm \\
    
\hline
    Refractive index of layer A & $n_A$ & 2.22 (Titanium Oxide) \\
    
\hline
    Width of layer A & $d_A$ & 71.26nm \\  
    
\hline
    Refractive index of layer B & $n_B$ & 1.41 (fused silica) \\ 
    
\hline
    Width of layer B & $d_B$ & 112.20nm \\
     \hline
    
\end{tabular}

\end{table}

\section{Results and Discussion}\label{sec3}
\subsection{Coherent control of defect mode properties in the PC}
\begin{figure}[htb]
\centerline{
\includegraphics[width=8.8cm]{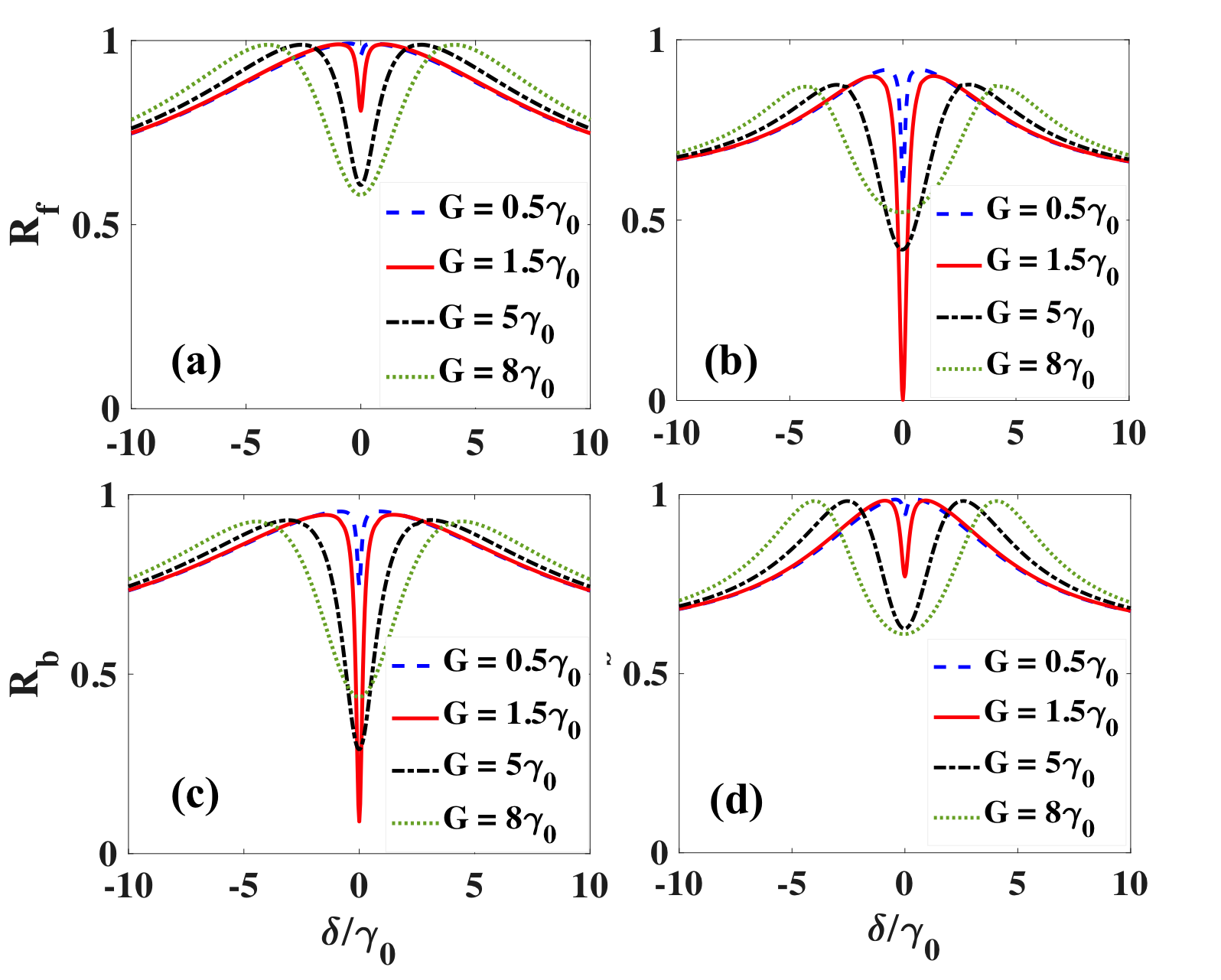}}
\caption{Reflection spectra [$R_{f,b}=|r_{f,b}|^2$] of the PC for different values of $G$ in (a),(b) forward direction and (c),(d) backward direction for the modes $\omega_1$  and $\omega_2$, respectively.}
\label{fig3}
\end{figure}

\begin{figure}[htb]
\centerline{
\includegraphics[width=8.8cm]{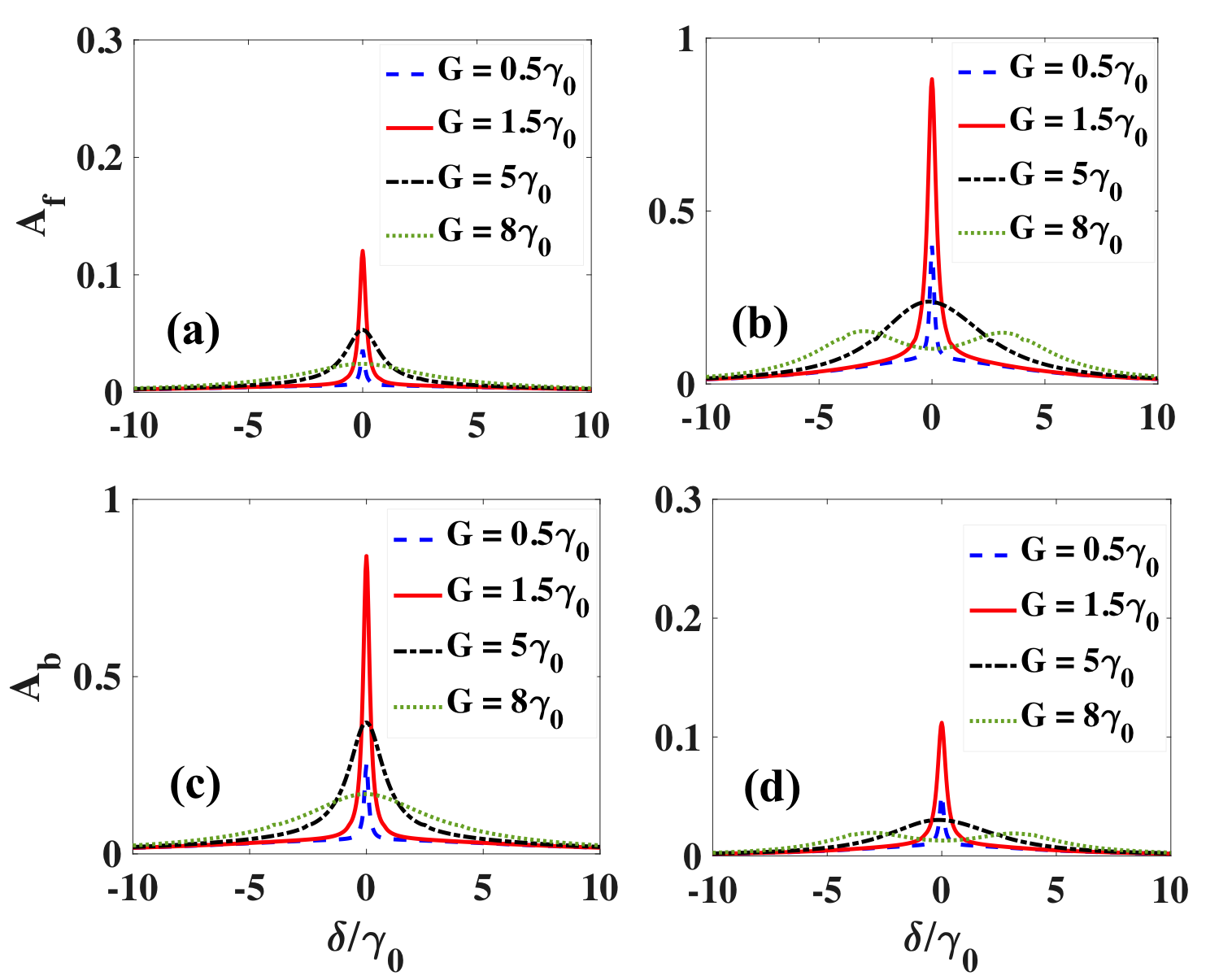}}
\caption{Absorption spectra [$A_{f,b}=1-(T_{f,b}+R_{f,b})$] of the PC for different values of $G$ in (a),(b) forward direction and (c),(d) backward direction for the modes $\omega_1$ and $\omega_2$, respectively.}
\label{fig4}
\end{figure}

As mentioned before, the PC  behaves as a two-mode cavity with two non-degenerate defect modes lying in the band gap region. We observe that these modes get red-shifted to the frequencies $\omega_1=2.61253 \times 10^{15}$ Hz and $\omega_2=3.329655 \times 10^{15}$ Hz, when the defect layer is doped with atoms. The parameters used for the layers in the PC are given in Table \ref{tbl1}. We first numerically obtain the spectra for transmission, reflection, and absorption, using the expressions of $t_f$, $t_b$, $r_f$, and $r_b$, and display them in Figs. \ref{fig2}, \ref{fig3}, and \ref{fig4}, respectively,  for both the forward (left-to-right) and the backward (right-to-left) direction of the input field, while it is near-resonant to either of the defect modes. We have used Eq. (\ref{subeq:3}) for susceptibility in our analysis, which enables us to incorporate the coherent control of spectral properties.


We first discuss the forward propagation of the field. It can be seen that at resonance, the transmission profile is nearly the same for both frequencies, irrespective of the directions of propagation [Figs. \ref{fig2}]. In the regime of weak control field ($G<\gamma_{13},\gamma_{23}$), the PC is nearly fully reflecting [Fig. \ref{fig3}a], and nearly non-absorptive [Fig. \ref{fig4}a], when the input field is resonant with the mode $\omega_1$. 
For larger Rabi frequencies, $G> (\gamma_{13}, \gamma_{23})$, the atoms get excited and the system starts showing a transmission peak at resonance, along with two new resonances for very large reflection and negligible absorption. Such resonance can be attributed to the Autler-Townes splitting of the excited state manifold in the three-level system. When the $G$ exceeds the Doppler width of the $|1\rangle \rightarrow |3\rangle$ transition, the transmission is increased to $35\%$ on resonance due to the saturation effect of dipoles, referring to dipole-induced transparency (DIT) \cite{PhysRevLett.96.153601}. Due to inhomogeneous broadening, arising out of the Doppler effect and the dephasing, complete transparency is not achieved. 
However, for values of $G$ smaller than the Doppler width, the PC remains predominantly reflecting, without much contribution from atoms \cite{PhysRevLett.64.1107} [Fig. \ref{fig2}]. 

Interestingly, for a critical intermediate value of $G$ ($=1.5\gamma_0$), the system exhibits maximum absorption $\sim 90\%$ [see Fig. \ref{fig4}(b)] and the minimum reflection (near to zero) at the mode frequency $\omega_2$ [Fig.\ref{fig3}(b)], implying an absorption saturation. There is an interplay in reflection and absorption in this regime and the dipole is said to be critically coupled to the defect mode $\omega_2$. The dip in the reflection at resonance can be attributed to a near-perfect destructive interference, as discussed in the context of non-linear medium as a spacer layer in a PC \cite{NireekshanReddy:13}. On the contrary, at the frequency $\omega_1$, the PC exhibits negligible absorption along with maximum reflection, without any signature of critical coupling. 

Hence, the optical properties of the PC in the presence of atomic gas can be tuned by the appropriate choice of the control field. 
The PC can act as a good transmitter for a strong control field and can be a good absorber or reflector when resonant to a mode frequency $\omega_2$ for the critical coupling. Also, the mode $\omega_1$ remains nearly lossless,  while the mode $\omega_2$ is lossy. 

\begin{figure}[htb]
\centerline{
\includegraphics[width=8.8cm]{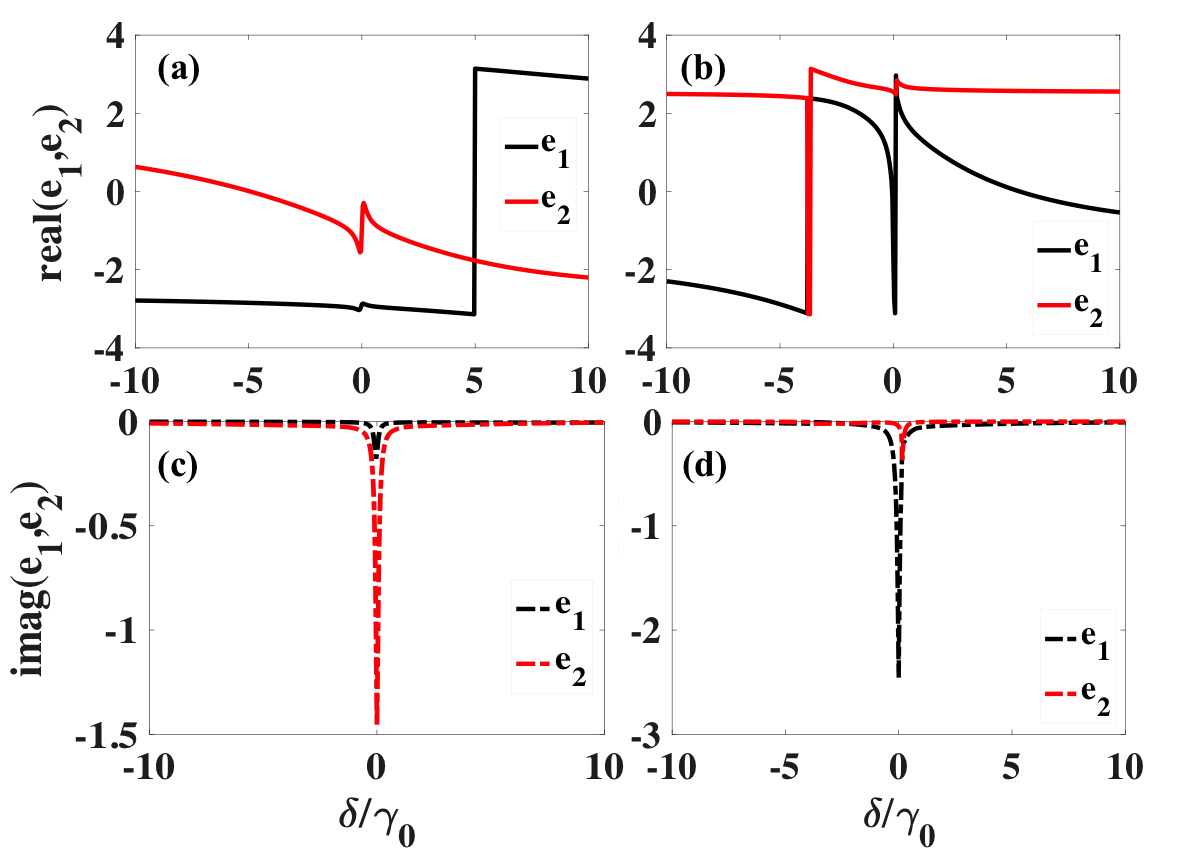}}
\caption{Spectra of real and imaginary components of eigenvalues ($e_1,e_2$) of Hamiltonian at $G=1.5\gamma_0$ for the defect modes $\omega_1$ [(a),(c)] and $\omega_2$ [(b),(d)]. The other parameters are the same as in Fig. \ref{fig2}.}
\label{fig5}
\end{figure}

\subsection{Directional response of spectral properties of defect modes}
We next investigate the directional response emerging in the PC due to the broken PT symmetry. We focus on the critical coupling condition. As can be seen from the Figs. \ref{fig3}(a) and (c), the reflection coefficient of the PC is quite different for forward and backward propagation, at the resonance $\omega_1$. It is nearly unity (nearly zero) for forward (backward) propagation. This is compensated by the absorption coefficient, which is negligible (nearly unity) for forward (backward) propagation [see Figs. \ref{fig4}(a) and (c)]. This directional dependence corresponds to a non-reciprocity in both reflection and absorption. Interestingly, for the mode $\omega_2$ as well, the nonreciprocity exists, but with opposite features: reflection becomes maximum for backward propagation [Fig. \ref{fig3}(d)], and the absorption maximum for forward propagation [Fig. \ref{fig4}(b)]. The nonreciprocal effect exhibited at $\omega_1$ with reference to the forward direction can be seen at $\omega_2$ when viewed backward. 

Here we must point out the distinctive features of our results. In previous works, by suitably modulating the gain/loss parameters in the presence of special types of singularities, one can lead to the CPA and lasing for the same \cite{PhysRevLett.113.263905} and the different wavevector amplitudes \cite{Hang_2016}. Our case does not correspond to CPA and lasing, as we do not work at zeroes of the diagonal elements of the matrix $M$. We strictly avoid any singularity, still achieving non-reciprocity. As we mentioned before, at both the resonances, the transmission however remains the same and independent of the direction of propagation. Although the transmission increases (up to $42\%$) in the system with higher values of the control field due to the saturation effect, the possibility of lasing modes is not seen.  

The above result can be interpreted in terms of the energy spectra of the PC. We display the real and imaginary parts of the Hamiltonian $H$ (see Sec. \ref{sec2}) in Figs. \ref{fig5}. It can be easily seen that the real values of the eigenvalues at both the resonances cross each other at a specific value of detuning (e.g., $\delta=5\gamma_0$ for mode 1 and $\delta=-4\gamma_0$ for mode 2), and at these level crossings (pertaining to a phase transition), the corresponding imaginary parts remain zero, such that the eigenvalues are real. This implies that Hamiltonian remains hermitian in this region. In fact, the Hamiltonian has a real spectrum except at the resonance $\delta = 0$, when the imaginary parts become nonzero and the Hamiltonian becomes non-hermitian. Moreover, the imaginary part of one of the eigenvalues is small (large) negative at resonance with $\omega_1$ ($\omega_2$) [black curves in Figs. \ref{fig5}(c) and (d)] and vice versa for the other eigenvalue (red curves in these figures). This suggests that while one of the modes will be highly reflective, the other will be highly absorptive. These modes are classified as non-PT dual pairs with one being a gain (amplifying) mode and the other being a lossy (dissipative) mode \cite{PhysRevA.87.012103}. In PT-symmetric systems, the gain and loss are balanced in a way that ensures real-valued eigenvalues. But in non-PT symmetric systems, this balance is disrupted, leading to complex eigenvalues that are not complex conjugate of each other (as depicted by the asymmetry in the plots of imaginary parts of the eigenvalues, in Figs. \ref{fig5}). In addition, eigenvalues in Figs. \ref{fig5} do not show the possibility of diagonal elements of transfer matrix $M_{11}(\omega)$ and $M_{22}(\omega)^*$ to be zero, which ensures the absence of singularity in the system. Further, we note that the eigenvectors of the system do not coalesce, assuring the absence of EPs in the system. Therefore, the presence of singularities is not essential to observe phase transitions and non-reciprocal reflections in non-PT-symmetric PCs unlike \cite{PhysRevLett.113.263905,Hang_2016,article14}.
\begin{figure}[htb]
\centerline{
\includegraphics[width=8.8cm]{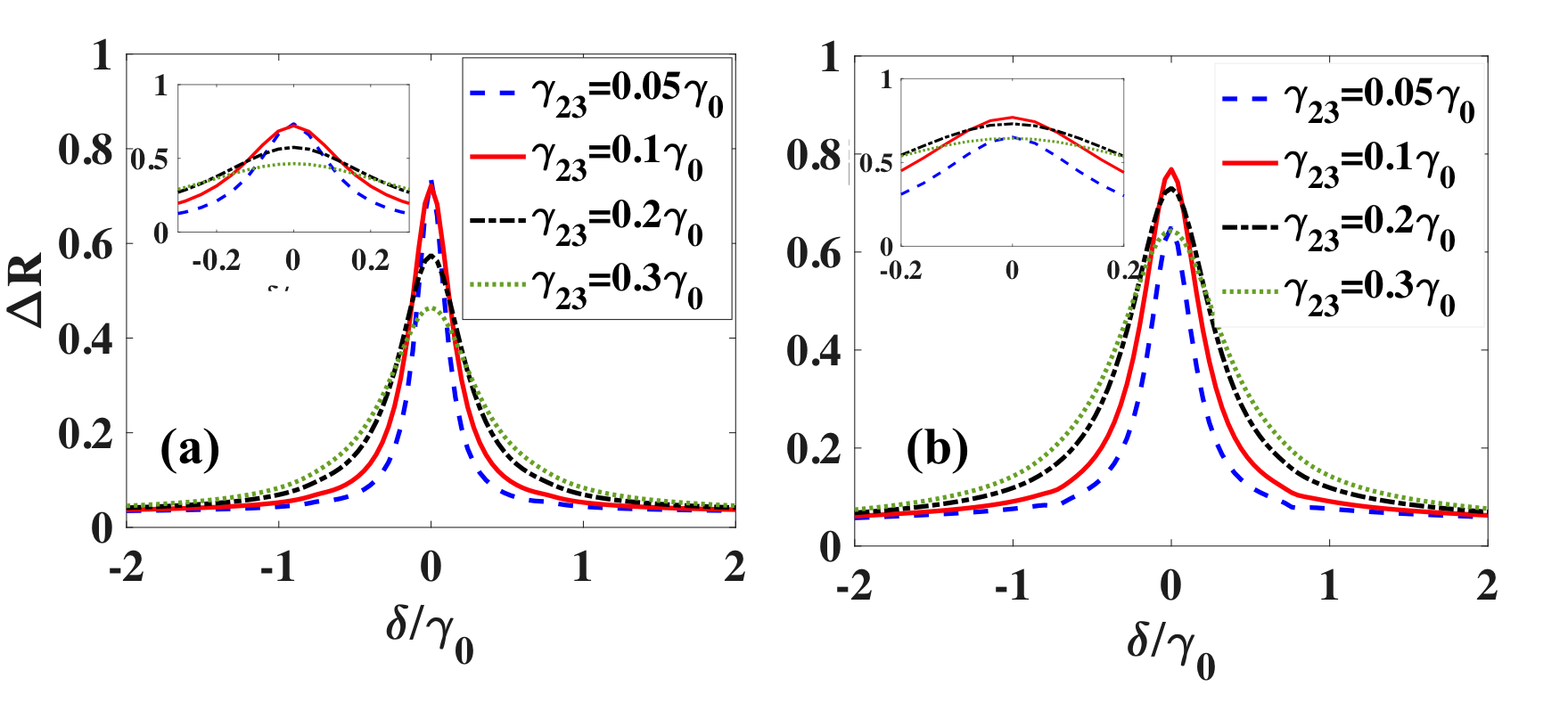}}
\caption{Variation in $\Delta R$ with detuning for different values of the decay rate $\gamma_{23}$, when $G = 1.5\gamma_0$, $\gamma_{13} = 0.1\gamma_0$, and the field is at resonance with (a) $\omega_1$ and (b) $\omega_2$. The other parameters are the same as in Fig. \ref{fig2}.}
\label{fig6}
\end{figure}

The nonreciprocity refers to a difference in the reflection coefficients, $\Delta R = R_f-R_b$ for forward and backward propagation, at resonance. For $G = 0.1\gamma_0$ (equal to the atomic decay rate), $\Delta R \approx 0$, referring to a nearly reciprocal reflection in the PC, when resonant to either mode (see Figs. \ref{fig3}). But, for larger $G$, $\Delta R$ increases and becomes maximum when $G$ takes the critical value $1.5\gamma_0$. With further increase in $G$, the nonreciprocity deteriorates, e.g., less than $20\%$ for $G=8\gamma_0$. This can be attributed to larger Rabi splitting for stronger coupling between the atoms and the defect mode, akin to the atom-cavity coupling reported in \cite{PhysRevLett.68.1132}.

Further, we show in Figs. \ref{fig6} the variation of the nonreciprocity $\Delta R$ with the detuning $\delta$ for different values of the spontaneous dissipation rate $\gamma_{23}$ of the atoms in transition $|2\rangle \rightarrow |3\rangle$, when the field is resonant to either mode. We find that for $\gamma_{23} = \gamma_{13}$, the $\Delta R$ is maximized and for larger $\gamma_{23}$, it decreases. This happens due to the imbalance of the population in the energy levels which affects both the absorption and reflection of the input intensity depending on the direction of the input field. While such dependence of non-reciprocity (in transmission) on atomic dissipation rates arises due to non-linearity and time reversal asymmetry in a cavity-QED setup \cite{Xia:14},  it is the absence of PT-symmetry in our case, that leads to nonreciprocity.
We further emphasize that similar results, as reported in this paper with three-level atomic configurations, have also been obtained with atoms having two-level configurations. Moreover, this study has also been verified if the substrate defect layer is vacuum and two-level and three-level atoms are doped on it. 

\section{Conclusion}
We have shown how a defect layer doped with three-level atoms in a one-dimensional multilayered PC can give rise to the breaking of PT-symmetry. This is in contrast to the previous studies, which considered only the bilayer systems with an imbalance in gain/loss parameters to break the PT-symmetry. In our case, the broken PT symmetry leads to nonreciprocity in reflection and absorption at two defect modes. This non-reciprocity can be controlled by the coherent field, that drives the atoms and the dissipation rates of the atoms. While at the critical coupling, the system at resonance with defect mode 1 acts as a nearly perfect absorber in the forward direction and a good reflector in the reverse direction, and vice-versa for defect mode 2. Hence, the role of defect modes gets interchanged on reversing the direction of the incident field and they are termed as non-PT dual pairs. Furthermore, the relevant Hamiltonian obtained from the scattering matrix does not exhibit any EP. This implies that our study reveals a new underlying physics that the presence of an EP is not necessary for nonreciprocity.

\appendix


 \bibliographystyle{elsarticle-num} 
 \bibliography{main}





\end{document}